\documentclass[12pt,floatfix,aps,nofootinbib,superscriptaddress]{revtex4} 
\pdfoutput=1 
\usepackage{amssymb,amsfonts,multirow}

\usepackage{youngtab}

\newcommand{\fund}{\tiny\Yvcentermath1\yng(1)}
\newcommand{\afund}{\tilde{\tiny\Yvcentermath1\yng(1)}}

\usepackage{overpic}
\usepackage{feynmp-auto} %creates FD

\usepackage{fix-cm}    

\makeatletter
\newcommand\HUGE{\@setfontsize\Huge{45}{100}}
\makeatother

\usepackage{epsfig}
\usepackage{bbm}

\usepackage{placeins}
\usepackage{xspace}
\usepackage{cancel} 
\usepackage{appendix}

\usepackage{amssymb,url}
\usepackage{graphicx}
\usepackage{hyperref}

\usepackage{color}

\usepackage{array}
\usepackage{amsmath}
\usepackage{slashed}

\definecolor{rossoCP3}{cmyk}{0,.88,.77,.40}

\long\def\del #1 \enddel { }

\usepackage{graphicx}
\usepackage{amsmath}
\usepackage{amssymb}
\usepackage{subfigure}

\usepackage{epsfig}

\usepackage{graphicx}
\usepackage{subfigure}
\usepackage{hyperref}

\def\beq{\begin{equation}}
\def\eeq{\end{equation}}

\def\bea{\arraycolsep .1em \begin{eqnarray}}
\def\eea{\end{eqnarray}}
\def\Tr{{\rm Tr}}

\def\eps{\epsilon}

\def\s0#1#2{\mbox{\small{$ \frac{#1}{#2} $}}}
\def\0#1#2{\frac{#1}{#2}}

\def\grgl{\:\hbox to -0.2pt{\lower2.5pt\hbox{$\sim$}\hss}{\raise3pt\hbox{$>$}}\:}
\def\klgl{\:\hbox to -0.2pt{\lower2.5pt\hbox{$\sim$}\hss}{\raise3pt\hbox{$<$}}\:}

\def\lsim{\mathrel{\rlap{\lower4pt\hbox{\hskip1pt$\sim$}}
    \raise1pt\hbox{$<$}}}                % less than or approx. symbol
\def\gsim{\mathrel{\rlap{\lower4pt\hbox{\hskip1pt$\sim$}}
    \raise1pt\hbox{$>$}}}                % greater than or approx. symbol

\begin{document}
${}$\vskip1cm

\title{Framework for an asymptotically safe Standard Model via dynamical breaking
${}$\vskip1cm
}
\author{Steven Abel}
\email{s.a.abel@durham.ac.uk}
\affiliation{\mbox{IPPP,
Durham University, South Road, Durham, DH1 3LE}}
\affiliation{\mbox{Cern, Theoretical Physics Department, 1211 Geneva 23, Switzerland}}
\author{Francesco~Sannino$^{2,\,}$}
\email{sannino@cp3-origins.net}
\affiliation{{\color{rossoCP3}CP${}^3$-Origins} \& the Danish Institute for Advanced Study,
Univ. of Southern Denmark, Campusvej 55, DK-5230 Odense}

\begin{abstract}
\vskip2cm
\noindent
We present a consistent embedding of the matter and gauge content of the  
Standard Model into an underlying asymptotically-safe theory, that has a well-determined  
interacting UV fixed point in the large colour/flavour limit. 
The  scales of symmetry breaking are determined by two  mass-squared parameters
with the breaking of electroweak symmetry being driven radiatively. 
There are no other free parameters in the theory apart from gauge couplings.   

\vskip7.cm
{\noindent \footnotesize Preprint:  CERN-TH-2017-156, CP3-Origins-2017-028 DNRF90, IPPP-2017/59}

\end{abstract}
\maketitle
\newpage
%\tableofcontents

\section{Introduction}
Recent work has demonstrated that a general class of asymptotically-safe gauge-Yukawa theories  \cite{Litim:2014uca,Antipin:2013pya,LMS,Intriligator:2015xxa,Bajc:2016efj,Pelaggi:2017wzr}  supports a form of radiative symmetry breaking reminiscent of that in the Minimal Supersymmetric Standard Model (MSSM)
\cite{Abel:2017ujy}. That is, even if Higgs mass-squareds are positive in the ultra-violet (UV), they are driven negative radiatively in the infra-red (IR)  by their Yukawa coupling to quarks. The presence of ``Higgs" scalars, in perturbation theory, that couple to quarks is a necessity for the theory to have a UV fixed point  \cite{Litim:2014uca}, so this form of radiative symmetry breaking is intimately connected with the  asymptotic safety of the theory. In the framework of asymptotic safety, such theories are {\it technically} natural in the 
sense that they are  determined simply by the choice of  renormalization group (RG) trajectory in
the space of all relevant operators (including mass-squareds).

An interesting outstanding question is then whether the Standard Model (SM) can be embedded in a natural way into such theories. 
This follow-up paper answers the question in the affirmative; it is demonstrated that, while not exactly trivial, a suitable embedding can be 
constructed in an extremely straightforward fashion. The resulting UV complete incarnation of the SM has no Landau poles for any couplings -- even hypercharge -- and is genuinely asymptotically safe\footnote{Modulo the eventual inclusion of gravity: it is beyond the scope of the present work to discuss gravity, but -- as with supersymmetry versus supergravity -- we adopt the approach that  asymptotic-safety of gravity can be established independently without disrupting the present discussion.}. 

\begin{figure}
\noindent \begin{centering}
\includegraphics[scale=0.95,bb=0bp 60bp 250bp 260bp,clip]{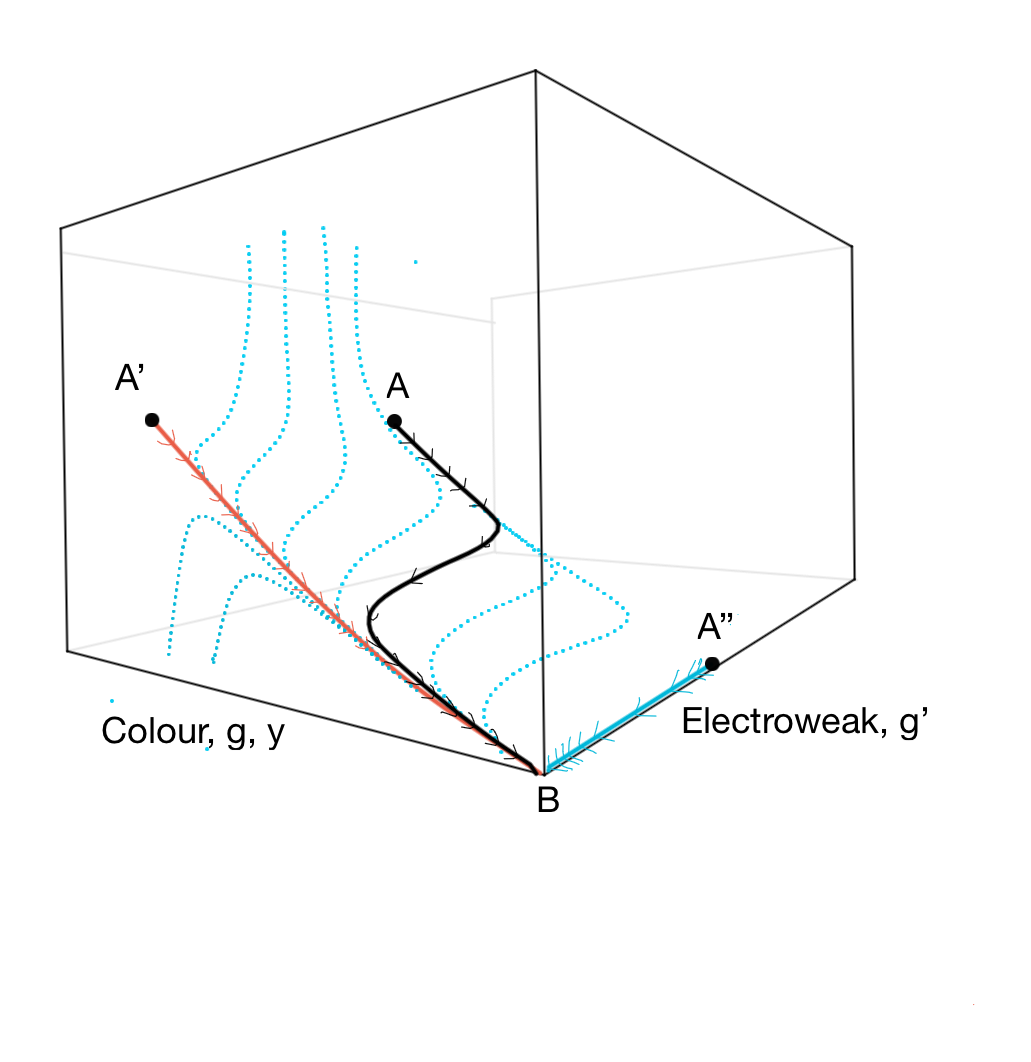}
\includegraphics[scale=0.4,bb=0bp 0bp 500bp 670bp,clip]{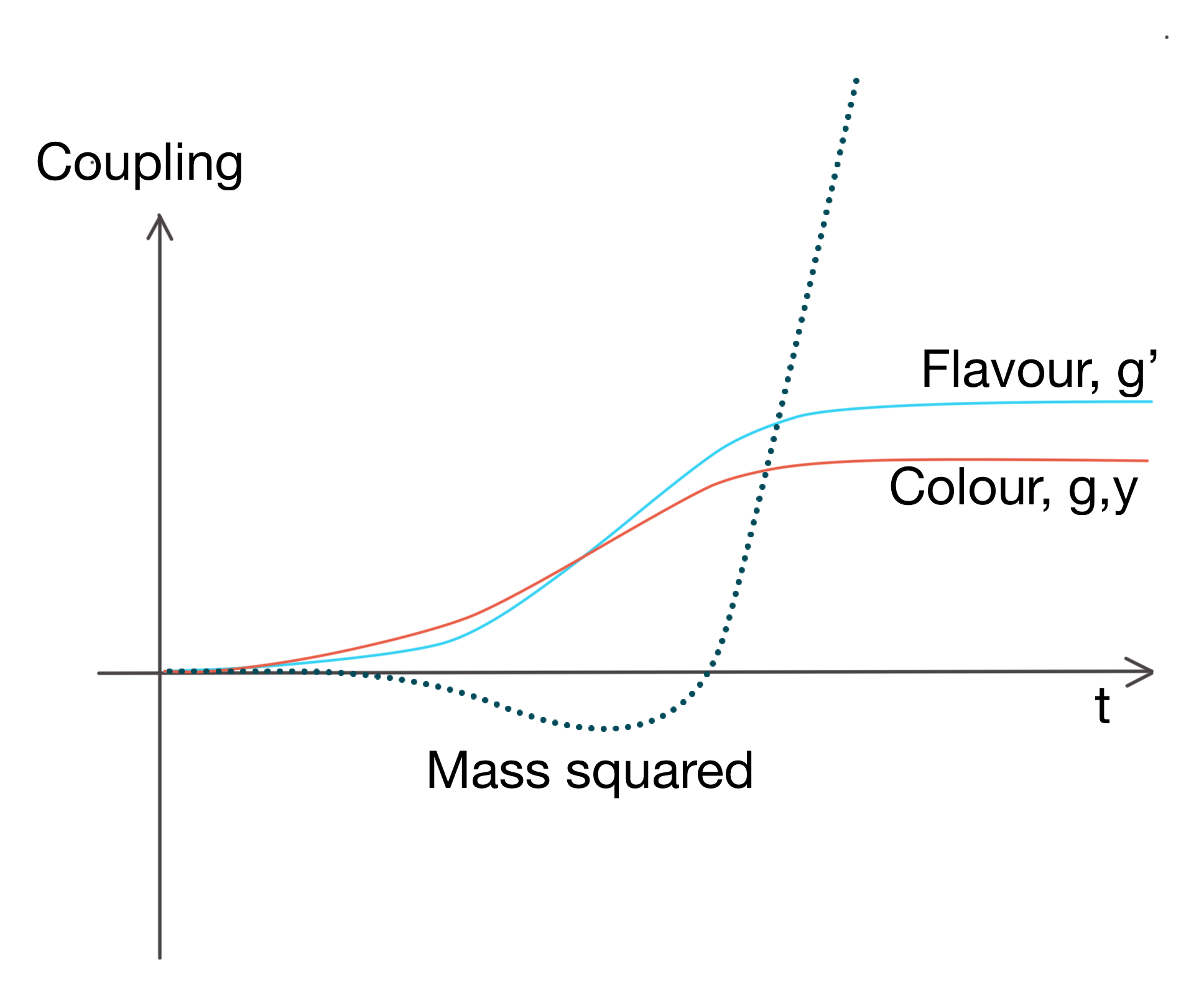}
\par\end{centering}
\protect\caption{The renormalisation group flow of the couplings from the UV fixed point and around the critical
curve, towards the Gaussian IR fixed point. The black line is fixed by matching the desired electroweak couplings at the low scale. \label{fig:flow}}
\end{figure}

The construction consists of an embedding of the SM into an extended Pati-Salam-like theory, whose gauge symmetry is spontaneously broken to the SM gauge group, $SU(N)\times SU(2)_L\times SU(2)_R  \rightarrow  SU(3)_c\times SU(2)_L\times U(1)_Y$. The flow in such a theory takes a very generic form, as displayed in figure \ref{fig:flow}, from UV fixed point A towards a Gaussian fixed point B. 
The left panel shows the running in the strong/electroweak coupling-space, where the ``strong'' couplings comprise the $SU(N)$ gauge coupling, $g$, the Yukawa coupling $y$ plus the quartic scalar couplings, and where  the ``weak'' gauge couplings are referred to generically as $g'$. Calculable asymptotic safety can be achieved in  the large colour and large flavour limit. 

The true UV fixed point $A$ is related to the fixed point $A'$ when one turns off the weak gauge couplings. The latter corresponds 
precisely to the behaviour  of the pure $SU(N)$ gauge-Yukawa theories of  \cite{Litim:2014uca}, which have large numbers of colours $N$  and flavours $N_F$ of fermionic ``quarks''  coupling to an $N_F\times N_F$ scalar. By making a judicious ``Veneziano limit'' choice of $N$ and $N_F$, fixed point $A'$  can be made arbitrarily weakly coupled even though it is still interacting. 
However the $SU(N)$ gauging also implies that the electroweak gauge couplings see many fundamental ``flavours'', 
so fixed point A is also related in an orthogonal direction to the ``many flavour" UV fixed point $A''$ of the $SU(2)_R\times SU(2)_R$~ gauge group, corresponding to the behaviour when one turns off the strong $SU(N)$ gauge coupling. 

Naturally, there are two components to the spontaneous symmetry breaking in such a scenario, one for the Pati-Salam breaking and one for the electroweak breaking. The former requires the addition of  $N-3$ coloured scalars to break $SU(N)$ to the $SU(3)_c$ of the SM via a ``rank-condition''. It is highly non-trivial that there still exists a UV fixed-point when one adds order $N$ coloured scalars to the theory of \cite{Litim:2014uca} so that, regardless of the size of the $SU(N)$ gauge group, such a breaking is always possible with only a quantitative change in the UV fixed point. The spontaneous breaking is driven by the addition of a relevant and negative mass-squared operator for the
scalars (which being a relevant operator is of course unable to disturb the asymptotically safe fixed point). 

The second part of the breaking relies on the observation of  \cite{Abel:2017ujy} that if a positive mass-squared operator for the scalars is added to the theory it is driven negative in the IR, resulting in radiative symmetry breaking, with the running terminating at some point { on its way towards fixed point B}. As mentioned, this mechanism is analogous to the radiative symmetry breaking in the MSSM \cite{Ibanez:1982fr}.

The flow is also shown in the right panel of the figure as a function of RG scale. The ``strong'' couplings are actually overtaken in the UV by the ``weak'' couplings, due to the fact that their fixed point relies on the resummation technique of \cite{Holdom,Pica:2010xq}. Thus, although the coupling itself is still weakly coupled in the UV, because of the proliferation of electroweak fundamentals the 't Hooft coupling of the electroweak factor in the theory is order unity. Therefore a crucial part of the discussion will be to show that the overall picture is indeed as shown in   figure \ref{fig:flow}, with the two kinds of asymptotic behaviour governing the overall flow, but not interfering. 

We should stress that there is very little freedom in the framework: the above description is simply what happens when coloured scalars 
are added to the theory in  \cite{Abel:2017ujy}, and an $SU(2)_L\times SU(2)_R$  subgroup of the global symmetry gauged. In particular there are very few arbitrary parameters. In fact there are only two free parameters besides the gauge couplings in the theory, which as usual correspond to the relevant operators, namely the mass-squareds. In accord with the most predictive  asymptotic safety picture, the asymptotically safe couplings are all fixed in terms of the gauge coupling along the flow, because the theory has a single trajectory between the two fixed points A and B. Meanwhile every relevant operator represents a new degree of freedom or, equivalently, a loss of predictivity. In the present case there is a free parameter for each  scale of symmetry breaking. Because the other couplings are all constrained, it is then non-trivial that the theory turns out to be stable (that is there are no negative quartic couplings along the flow). In addition portal-couplings between the coloured scalars and the Higgses turn out to be IR-free. Hence they cannot disrupt the flow but (like other irrelevant operators) are precisely zero at the UV fixed point, in accord with the standard asymptotic safety prescription. 

The layout of the discussion is as follows: the next section recaps the structure of the UV complete theories of \cite{Litim:2014uca,Antipin:2013pya,LMS,Intriligator:2015xxa,Bajc:2016efj,Pelaggi:2017wzr} which form the core of the UV fixed point theory, and then indicates how the SM can be embedded into it, and the additional states that must be added. Here the focus is on the general structure which is as we have said broadly speaking a many colour/flavour extension of the Pati-Salam model. We discuss the symmetry breaking pattern, where the gauge groups and matter fields of the SM fit, and which are the crucial operators.  Section \ref{sec3} then goes on to discuss the RG flow, and establishes that the symmetry breaking does in fact occur in the desired way. In particular it demonstrates that the additional coloured scalars in the theory do not destroy the UV fixed point, and that at least one UV fixed point gives rise to a stable flow.

Finally we discuss the running of the electroweak $SU(2)_L\times SU(2)_R$ groups of the SM. One has to establish that this gauging {\it also} does not disrupt the original fixed point, and that the electroweak gauge couplings run independently to their own fixed points in the UV. This part of the discussion utilises a straightforward adaptation of the ``large number of flavours'' limit of \cite{Holdom,Pica:2010xq} (where ``flavours'' in this case means fundamentals of the electroweak gauge group); but the important point is that this ingredient can be added in an independent modular fashion because, as we demonstrate, the electroweak running and the running in the core $SU(N)$ theory decouple in the Veneziano limit.  

We should stress that there are most likely many other configurations within the general framework, and this paper presents the most minimal realisation of asymptotic safety in the SM via radiative breaking. Moreover we view this framework as just a first step towards an asymptotically safe SM of this kind. A more complete treatment must address for example flavour structure and fermion mass hierarchies, which we do not treat in detail here. 
  
\section{Embedding the SM} 

We first focus on the structure of the SM embedding, building up from the theories discussed in  \cite{Litim:2014uca,Antipin:2013pya,LMS,Intriligator:2015xxa,Bajc:2016efj,Pelaggi:2017wzr,Abel:2017ujy}. These are $SU(N)$ gauge theories with  $N_F$ flavours of fermion pairs $Q^i_L$,~${Q}^i_R$ $(i=1,\cdots,N_F)$ in the fundamental representation, and an $N_F\times N_F$ complex matrix scalar field $H$ uncharged under the $SU(N)$ gauge group. The particle content is shown in Table \ref{sqcd0-2-1}.

\begin{table}
\centering{}%
\begin{tabular}{|c||c||c|c|}
\hline 
$ $ & $SU(N)$ & $SU(N_{F})_{L}$ & $SU(N_{F})_{R}$\tabularnewline
\hline 
\hline 
$Q_{L,a}^{i}$ & $\fund$ & $\afund$ & $1$\tabularnewline
\hline 
$Q_{R,i}^{a}$ & $\afund$ & $1$ & $\fund$\tabularnewline
\hline 
$H_{i}^{j}$ & $1$ & $\fund$ & $\afund$\tabularnewline
\hline 
\end{tabular}\caption{\emph{Fields in the basic model} of \cite{Litim:2014uca}\label{sqcd0-2-1}.}
\end{table}

We will throughout closely follow the original notation in  \cite{Litim:2014uca}, 
using $i,j,k...$ to label flavour, and $a,b,c$... to label colour.
The Lagrangian is given by the sum of the Yang-Mills term, the fermion and scalar kinetic terms, the Yukawa interaction, and scalar self-interaction terms:
\begin{align}
\label{F2}
\mathcal{L}_{UVFP} & ~=~ \mathcal{L}_{YM}+\mathcal{L}_{KE}+\frac{y}{\sqrt{2}}\text{Tr}\left[\left(Q^\dagger_{L} H \cdot Q_{R}\right)\right]-u_{1}\text{Tr}\left[H^{\dagger}H\right]^{2}-u_{2}\text{Tr}\left[H^{\dagger}H\,H^{\dagger}H\right] ~ ,
\end{align}
where the decomposition $Q=Q_L+Q_R$ with $Q_{L/R}=\frac 12(1\pm \gamma_5)Q$ is understood. The trace $\Tr$ indicates the trace over flavour indices while the dot-product refers to $SU(N)$ colour. Ref.\cite{Litim:2014uca} discovered a number of UV fixed points for this model in the Veneziano limit where 
$N\gg 1$ with  
\begin{equation}\label{eps}
\eps~=~\frac{N_F}{N}-\frac{11}{2} ~\ll ~1~ .
\end{equation}
As we will later see, in this limit the 't Hooft couplings are all proportional to the parameter $\epsilon$, which is therefore an indicator of the perturbative reliability of the fixed point.

\begin{table}
\centering{}%
\begin{tabular}{|c||c||c|c|c|}
\hline 
$ $ & $SU(N)$ & $SU(N_{F})_{L}\supset SU(2)^{}_{L}$ & $SU(N_{F})_{R}\supset SU(2)^{}_{R}$ & $SU(N_S) \supset SU(2)_{R}$\tabularnewline
\hline 
\hline 
$Q_{L,a}^{i}$ & $\fund$ & $\afund$ & $1$ &$ 1 $\tabularnewline
\hline 
$Q_{R,i}^{a}$ & $\afund$ & $1$ & $\fund$ & $1 $\tabularnewline
\hline 
$H_{i}^{j}$ & $1$ & $\fund$ & $\afund$ & $1$\tabularnewline
\hline 
$\tilde{Q}_{j=1..N_{S}}$ & $\afund$ & $1$ & $1$ & $\fund$ \tabularnewline
\hline 
\end{tabular}\caption{\emph{Fields in the extended model, where $\tilde{Q}$ are scalars and in the simplest case $N_S=N-3$. 
The $SU(2)_L\times SU(2)_R$ global flavour subgroups are then gauged, but note that three generations of fermions (i.e. the first 6 entries in $N_F$) transform under them. There are correspondingly 9 Higgs pairs: for the generations, one could identify an $SU(3)_L\times SU(3)_R$ SM flavour subgroup of the $SU(N_F)_L\times SU(N_F)_R$ symmetry, but to avoid complication we take this as implicit. The first two flavours of $\tilde{Q}$ form a fundamental of $SU(2)_R$ so that the  $SU(4)\times  SU(2)_L\times SU(2)_R$ Pati-Salam subgroup is broken down to the SM in the usual manner (namely with the $\tilde{Q}\supset (\overline{\bf 4}, {\bf 1},{\bf 2}) $ being the canonical Pati-Salam Higgs).  }\label{sqcd0-2}}
\end{table}

For the moment  let us focus on the embedding of the SM, which is shown in Table \ref{sqcd0-2}.
Our approach will be to embed $SU(3)_c$ of the SM into the 
$SU(N)$ of this theory. Therefore the first extra ingredient in the Table is $N-3$ scalar fundamentals of $SU(N)$, which we refer to as $\tilde{Q}$. Note that this is just one possibility for a breaking pattern which happens to be the simplest. We will somewhat reluctantly refer to these objects as squarks.  
 One can indeed take some lessons from supersymmetry regarding their possible properties, for example the fact 
  that one can add into the theory a positive mass-squared for them which (since they do not have Yukawa couplings to fermions) 
  will remain positive throughout the flow. 
  In order to arrive at the SM, the squarks will acquire VEVs
in the IR, breaking $SU(N)\rightarrow SU(3)_c$, and therefore one of the two necessary relevant operators that we add into the theory 
is a negative mass-squared for them.

By making suitable colour and flavour rotations they can be written in the 
form  
\begin{align*}
\langle \tilde{Q} \rangle \,\, 
& =\,\, \overbrace{\left( \begin{array}{cccccccccc}
0 & 0 & 0 ~& 1 & && & & &  \\
~ \vdots & \vdots & \vdots ~& && &\ddots &&&  \\
 0 & 0 & 0 & ~& &&&  && 1 
\end{array}\right) }^{N} \,.  
\end{align*}

The breaking induced on the colour side is 
\begin{equation} 
[ SU(N) ] \times SU(N-3) \rightarrow [ SU(3)_c ] \times SU(N-3)_{Diag}\, ,
\end{equation}
where the square brackets indicate that the symmetry is gauged, while the $SU(N-3)_{Diag}$ symmetry is the squark flavour symmetry. Ultimately an $SU(2)_R$ subgroup of this symmetry is identified with the electroweak factor, so that the $\tilde{Q}$ VEV  results in the standard Pati-Salam breaking. Counting degrees of freedom, all $ N^2-9$ Goldstone modes of the symmetry breaking are eaten by gauge bosons. 
There are then $2N (N-3) - (N^2-9) = (N-3)^2$ real degrees of freedom remaining from the 
$\tilde{Q}$, which are all ``Higgses'' with masses of order the breaking scale.

  Secondly in the Table, we  indicate the assignment of the states and  the embedding of the global symmetries. 
  The assignment of SM matter fermions inside $Q_{L/R}$ is as three generations of $SU(2)_{L/R}$ doublets in the first 6 flavour entries, as shown explicitly  in  (\ref{eq:ql}). 
  The horizontal dots indicate states that are necessarily also charged under the $SU(2)_L\times SU(2)_R$ symmetry (as each row forms a single $SU(N)\times SU(2)_{L/R}$ bi-fundamental). As these entries are all charged under the broken part of the $SU(N)$ symmetry they cannot be produced in colliders at energy scales below the $SU(N)$ breaking scale. 
  The lower rows are singlets of $SU(2)_L\times SU(2)_R$ but are of course  still charged under colour. However this part of the theory 
  is non-chiral so for consistent phenomenology one is free to  add $m_Q {\bar Q}_L Q_R$  mass terms for all of these flavours (which would obviously change the flow at energy scales below $m_Q$). 
  Note that the leptons appear as the 4$^{th}$ colour, so the theory is indeed 
  effectively an $SU(N)\times SU(2)_L\times SU(2)_R$ extension of the Pati-Salam model. 
  
  Given this assignment of matter fields, and the Yukawa coupling in (\ref{F2}),  
  the first $6\times 6$ block of $H$ must fall into 9 bi-fundamentals of 
  $SU(2)_L\times SU(2)_R$ as shown explicitly in (\ref{eq:higgs}). (Note that $SU(2)$ contraction in the Lagrangian is with $\varepsilon = i\sigma_2$ tensors). 
  These 18 Higgs doublets would be the only possible source for  generating the flavour structure  in the effective quark Yukawas, so clearly in a fully phenomenologically viable model one would want the VEV of $H$  to be dominated by $H_{66}$. For the discussion in this paper we shall for simplicity maintain flavour symmetry, so the VEV for $H$ will be degenerate in the diagonal entries. Of course the entire first six columns (respectively rows) of $H$ fall into $SU(2)_L$ (respectively $SU(2)_R$) doublets.    
  
  The scalar field, $\tilde {Q}$ has only two flavours charged under $SU(2)_R$. In order to achieve the correct breaking down to the SM it is of course the uncharged field $\tilde{\nu}_R^e$ that gets a VEV along with the other $N-4$  fields along the top row of $\tilde{Q}$ that are uncharged under the SM gauge group.
  
   \begin{equation}
 \label{eq:ql}
Q_{L}~=~\left(\begin{array}{ccc}
q_{1} & \ell_{1} & \cdots\\
q_{2} & \ell_{2} & \cdots\\
q_{3} & \ell_{3} & \cdots\\
\vdots & \vdots & \ddots
\end{array}\right)\,;\,\,\,\,Q_{R}~=~\left(\begin{array}{ccc}
\left(\begin{array}{c}
u_{R}\\
d_{R}
\end{array}\right) & \left(\begin{array}{c}
\nu_{R}^{e}\\
e_{R}
\end{array}\right) & \cdots\\
\left(\begin{array}{c}
c_{R}\\
s_{R}
\end{array}\right) & \left(\begin{array}{c}
\nu_{R}^{\mu}\\
\mu_{R}
\end{array}\right) & \cdots\\
\left(\begin{array}{c}
t_{R}\\
b_{R}
\end{array}\right) & \left(\begin{array}{c}
\nu_{R}^{\tau}\\
\tau_{R}
\end{array}\right) & \cdots\\
\vdots & \vdots & \ddots
\end{array}\right)
\end{equation}
\begin{equation}
\label{eq:higgs}
H~=~\left(\begin{array}{cccc}
\left(\begin{array}{cc}
h_{d}^{0} & h_{d}^{-}\\
h_{u}^{+} & h_{u}^{0}
\end{array}\right)_{11} & \left(\begin{array}{cc}
h_{d}^{0} & h_{d}^{-}\\
h_{u}^{+} & h_{u}^{0}
\end{array}\right)_{12} & \left(\begin{array}{cc}
h_{d}^{0} & h_{d}^{-}\\
h_{u}^{+} & h_{u}^{0}
\end{array}\right)_{13} & \cdots\\
\left(\begin{array}{cc}
h_{d}^{0} & h_{d}^{-}\\
h_{u}^{+} & h_{u}^{0}
\end{array}\right)_{21} & \left(\begin{array}{cc}
h_{d}^{0} & h_{d}^{-}\\
h_{u}^{+} & h_{u}^{0}
\end{array}\right)_{22} & \left(\begin{array}{cc}
h_{d}^{0} & h_{d}^{-}\\
h_{u}^{+} & h_{u}^{0}
\end{array}\right)_{23} & \cdots\\
\left(\begin{array}{cc}
h_{d}^{0} & h_{d}^{-}\\
h_{u}^{+} & h_{u}^{0}
\end{array}\right)_{31} & \left(\begin{array}{cc}
h_{d}^{0} & h_{d}^{-}\\
h_{u}^{+} & h_{u}^{0}
\end{array}\right)_{32} & \left(\begin{array}{cc}
h_{d}^{0} & h_{d}^{-}\\
h_{u}^{+} & h_{u}^{0}
\end{array}\right)_{33} & \cdots\\
\vdots & \vdots & \vdots & \ddots
\end{array}\right); ~~
\tilde{Q}~=~\left(\begin{array}{ccc}
\left(\begin{array}{c}
\tilde{u}_{R}\\
\tilde{d}_{R}
\end{array}\right) & \left(\begin{array}{c}
\tilde{\nu}_{R}^{e}\\
\tilde{e}_{R}
\end{array}\right) & \cdots\\
\vdots & \vdots & \ddots
\end{array}\right)
\end{equation}
Finally we must extend the couplings in the theory to incorporate the new scalars:
\begin{align}
\label{F2P}
\mathcal{L}_{UVFP} & = \mathcal{L}_{YM}+\mathcal{L}_{KE}+\frac{y}{\sqrt{2}}\text{Tr}\left[\left(Q^\dagger_{L} H \cdot Q_{R}\right)\right]-u_{1}\text{Tr}\left[H^{\dagger}H\right]^{2}-u_{2}\text{Tr}\left[H^{\dagger}H\,H^{\dagger}H\right]\nonumber \\
 & \,\,\,\,\,\,\,\, ~~~~~~\hspace{4cm} -w_{1}\,\text{Tr}[\tilde{Q}^{\dagger}\cdot\tilde{Q}]^{2}-w_{2}\,\text{Tr}[\tilde{Q}^{\dagger}\cdot\tilde{Q}\,\tilde{Q}^{\dagger}\cdot\tilde{Q}]~ ,
\end{align}
where the dots indicate colour contraction, and we reiterate that $SU(2)$ contractions are with $SU(2)$ tensors. 
The $u_1\,,u_2\,,w_1\,,w_2$ couplings provide stability. These should render an overall  positive  quartic coupling, but as they are asymptotically safe (i.e. take a non-zero value at the fixed point) this is out of our control: it will turn out to be a successful prediction of the fixed point that this is the case. 

Note that we do not consider the couplings 
\begin{equation}
{\cal{ L}} ~\supset~ -v_{1}\text{Tr}\left[H^{\dagger}H\right]\text{Tr}[\tilde{Q}^{\dagger}\cdot\tilde{Q}] - v_2 \Tr[H^\dagger H \, \tilde{Q}^\dagger \tilde{Q}]~.
\end{equation}
Such ``portal'' couplings are in principle  rather interesting as they would generate a mass-squared term for the electroweak 
Higgs from the Pati-Salam breaking. However in the context of asymptotic-safety this is only a possibility if the couplings turn out to be (marginally) relevant. If this were the case then one could set them to be on an RG trajectory where they grow from zero in the UV, to be significant in the IR. It turns out that for the theory at hand this does not happen; whenever the overall quartic couplings are positive at the UV fixed point, so that the theory is stable, these couplings turn out to be marginally irrelevant. This means they grow in the UV, and thus, like all other irrelevant couplings, must be set to be precisely to zero at the fixed point, where they will  remain all along the flow. Again the signs of the beta-functions of $v_{1,2}$ are beyond our control, which is another indication of  the predictivity of the framework.   
(There could of course exist some other theory with non-zero portal couplings that flows close to the UV fixed point of this one, 
but such a theory would not be UV complete, so could not  be considered asymptotically safe.) 

For this discussion we are maintaining the remaining flavour symmetry for simplicity. Therefore we do not for example add any 
flavour dependent couplings for the squarks, but it would be straightforward to extend the discussion to incorporate them.
Note that the above set of couplings is closed under renormalisation.

%\subsection{Triggering Pati-Salam and electroweak symmetry breaking with a mass-squared operator}

The desired pattern of breaking will be driven by the 
relevant mass-squared operators  
\begin{align}
\label{mass-squared}
\mathcal{L} & ~=~\mathcal{L}_{UVFP} - m_{\tilde{Q}}^2\, \text{Tr}[\tilde{Q}^{\dagger}\cdot\tilde{Q}] - {m_0^2} \,\Tr (HH^\dagger) + {\Delta^2} \sum_{a=1}^{N_F^2-1} \Tr(HT^a)\Tr(H^\dagger T^a) ~ .
\end{align}
The mass-squared operators for $H$ are the same as those considered in  \cite{Abel:2017ujy}, and as established in \cite{Abel:2017ujy} such operators can be driven negative radiatively, even if they are positive in the UV. In particular the diagonal component of $H$, i.e. Tr$H$, gets a negative mass-squared dominated by the {\it non-degeneracy} parameter, $\Delta^2$. The advantage of radiative breaking for the electroweak sector is that it allows a hierarchy between the Pati-Salam and weak scales, even if the mass-squared operators are all similar in size. 

By contrast $m_{\tilde{Q}}^2$ cannot get a large 
negative contribution radiatively because it does not give mass to any fermions through Yukawa couplings. Therefore it is chosen to be negative all along the flow.

As per the rest of the theory we are maintaining an $[ SU(N_F)_L\times SU(N_F)_R]_{diag}$   flavour symmetry in the mass-squared operators that in a more comprehensive treatment could easily be broken. We should add that of course the gauging of $SU(2)_L\times SU(2)_R$ itself breaks the flavour symmetry. For simplicity we will neglect this in the running of the mass-squareds. One can  confirm that it is somewhat smaller effect. The relevant criterion is the relative sizes of the contributions to the Higgs anomalous dimensions from the Yukawas and the $SU(2)_L\times SU(2)_R$ gauge couplings. These  are comparable to those in the SM itself so it is equivalent to neglecting electroweak gauge couplings in the SM and maintaining only the top-Yukawa in the running. A better quantitative treatment would include them, but adding flavour dependence makes the discussion very intricate.

\section{The perturbative UV fixed point, $A'$, with zero electroweak gauge coupling}

\label{sec3}

We shall in the next section establish the decoupling of the strong and the electroweak fixed points in the  large colour/flavour  limit. Therefore it is useful to, in this section, first establish the existence of a fixed point in
the absence of electroweak gauging, namely UV fixed point  $A'$. 

The UV fixed point may be made  perturbative in a particular (Banks-Zaks \cite{BZ}) 
 limit, and can be determined in much the same way as in the 
original model of \cite{Litim:2014uca,Antipin:2013pya,LMS,Intriligator:2015xxa,Bajc:2016efj,Pelaggi:2017wzr}. The theory without $\tilde{Q}$ 
scalars had a UV fixed point in the limit of large $N_F$ and $N$, with $N_F/N =  22/4+\epsilon$, where $\epsilon \ll 1$. In this  limit, the one-loop contribution to the beta function is order $\epsilon$, which allows it to be balanced against the 
two-loop contribution while remaining arbitrarily perturbative. (Indeed all the couplings are proportional to $\epsilon $ at the fixed point, and 
 as shown in  \cite{Litim:2014uca} the radius of convergence is $\epsilon =0.117$.)

The fact that this leads to a  UV rather than an IR fixed point (which would correspond to the conventional Banks-Zaks fixed point  \cite{BZ}) has to do with the {\it sign} of the two-loop contribution to the gauge beta-function, which receives a negative contribution from  the Yukawa coupling $y$. Thus the scalar $H$ plays a crucial role. 

In the present context we can employ the same procedure except now of course the $N_S$ scalars $\tilde{Q}$ also contribute to the gauge beta-function.  Let us describe  the determination of the fixed point in detail. 
We define
\begin{equation}
x_{F}=\frac{N_{F}}{N}\,\,;\,\,\,x_{S}=\frac{N_{S}}{N}=\,22-4x_{F}+4\epsilon\,,
\end{equation}
where again the one-loop gauge beta function is proportional to $\epsilon$. We will consider various possible values of $N_S$.
It is also convenient
to define the following rescaled couplings:
\begin{align}
\alpha_{g}&=\frac{Ng^{2}}{(4\pi)^{2}}\,;\,\,\alpha_{y}=\frac{Ng^{2}}{(4\pi)^{2}}\,;\,\,\alpha_{u_{1}}=\frac{N_{F}^{2}u_{1}}{(4\pi)^{2}}\,;\,\,\alpha_{u_{2}}=\frac{N_{F}u_{2}}{(4\pi)^{2}}\,; \\ & \,\,\alpha_{v_{1}}=\frac{N_{F}Nv_{1}}{(4\pi)^{2}}\,;\,\,\alpha_{w_{1}}=\frac{N^{2}w_{1}}{(4\pi)^{2}}\,;\,\,\alpha_{w_{2}}=\frac{Nw_{2}}{(4\pi)^{2}}~,
\end{align}
with the numerical scaling counting the multiplicity that the traces provide to the coupling.

In the Veneziano limit, the running of the gauge coupling is  slower by a factor $\epsilon$ than that of the other couplings. 
Therefore the general picture is,  as depicted in figure \ref{fig:flow}, one in which the theory if started at an arbitrary point in 
coupling-space (but with zero electroweak coupling) runs rapidly to the red critical line, and then crawls towards the Gaussian fixed point in the IR. In the context of asymptotic safety of course the flow is always along the critical line, emanating {\it precisely} from the UV fixed point. Therefore one can determine the value of all the couplings in terms of $\alpha_g$  not just at the fixed point itself but along the critical line, with the value of $\alpha_g$ substituting for RG scale.    

Therefore to determine the fixed points, the RG equations are required to order
$\alpha^{3}\equiv\epsilon\alpha^{2}$ in $\beta_{g}$ and to order $\alpha^{2}\equiv\epsilon\alpha$
in the other couplings. (Note that to reduce clutter we will throughout use the subscripts on the beta function to refer to the rescaled 
coupling, i.e. $\beta_g \equiv d\alpha_g/dt$). Let us begin with $\alpha_g$ and $\alpha_y$. Their beta functions are 
\begin{align}
\beta_{g} & ~=~\alpha_{g}^{2}\left(\frac{4}{3}\epsilon+(36-2x_{F})\alpha_{g}-x_{F}^{2}\alpha_{y}\right)~,\nonumber \\
\beta_{y} & ~=~\alpha_{y}\left(-6\alpha_{g}+(1+x_{F})\alpha_{y}\right)~ .
\end{align}
It is useful to focus on two special limits of $N_S$ and $N_F$, namely $x_S\rightarrow 0$ and $x_{F}\rightarrow22/4$,  and alternatively $x_S\rightarrow 1$ and $x_{F}\rightarrow21/4$
. The former case is when there is a finite constant number of scalars, where  in the
Veneziano limit they will have negligible impact on the fixed point
behaviour: it will provide a useful check against the previous results
in the model without scalars. The latter case is when there are order
$N$ scalars in the Veneziano limit: this is the case of interest
in the present context given the SM embedding of the previous section which requires $N_S=N-3$. 

The exact trajectory in these two limits is along 
\begin{equation}
\alpha_{y}\,~=~\,\frac{6}{1+x_{F}}\alpha_{g}\,~~\longrightarrow~\begin{cases}
\frac{12}{13}\alpha_{g} & :\,x_{F}\rightarrow22/4\\
\frac{24}{25}\alpha_{g} & :\,x_{F}\rightarrow21/4
\end{cases}\, ~.
\end{equation}
The arrows indicate taking the Veneziano limit, so that corrections of order $\epsilon$ should be understood. 
It is convenient to define a parameter
$\sigma$ : 
\begin{equation}
0\,<\,\sigma~=~4x_{F}^{2}-17x_{F}-18\,~~\longrightarrow~\begin{cases}
\frac{19}{2} & :\,x_{F}\rightarrow22/4\\
3 & :\,x_{F}\rightarrow21/4
\end{cases}~.
\end{equation}
The fixed point in the gauge coupling is then found to be at 
\begin{equation}
\label{eq:astar}
\alpha_{g}^{*}~=~\frac{2}{3}\frac{(1+x_{F})}{\sigma}\epsilon\,~~\longrightarrow~\begin{cases}
\frac{26}{57}\epsilon & :\,x_{F}\rightarrow22/4\\
\frac{25}{18}\epsilon & :\,x_{F}\rightarrow21/4
\end{cases} ~.
\end{equation}
The constraint $\sigma>0$ comes from the requirement of positive
$\alpha_{g}^{*}$, and it translates into $x_{F}>\frac{17+\sqrt{577}}{8}=5.13$
and hence $x_{S}<1.49,$ which is of course compatible with both $x_{F}\rightarrow22/4+\epsilon$
with $x_{S}\rightarrow0$, and $x_{F}\rightarrow21/4+\epsilon$ with
$x_{S}\rightarrow1$. (Note that a similar calculation in a supersymmetric theory finds too many scalars in accord with \cite{Intriligator:2015xxa}, so it is
non-trivial that there are apparently order $N$  theories with fewer scalars that do have a solution.)
We will solve for $\alpha_g(t)$ where $t=\log(\mu/\mu_0)$ at the end.

Next we turn to $\beta_{u_{2}},\beta_{w_{2}}$ as these RG equations
involve only the couplings themselves and $\alpha_{y}$ and $\alpha_{g}$
respectively. As for the $y$ coupling, we need only keep terms up
to $\alpha^{2}$, and there is no suppression in the beta function,
so the flow is fast: 
\begin{align}
\beta_{u_{2}} & ~=~8\alpha_{u_{2}}^{2}+2\alpha_{u_{2}}\alpha_{y}-\frac{x_{F}}{2}\alpha_{y}^{2}\,~,\nonumber \\
\beta_{w_{2}} & ~=~32(6-x_{F})\alpha_{w_{2}}^{2}-6\alpha_{w_{2}}\alpha_{g}+\frac{3}{16}\alpha_{g}^{2}\,~.
\end{align}
One can see that the critical line is attractive to these couplings
because both $\alpha_{g}$ and $\alpha_{y}$ flow to zero in the IR,
and in this limit the above beta functions are both positive (provided
$x_{F}<6$) regardless of the sign of $\alpha_{u_{2}},\alpha_{w_{2}}$.
Solving for the beta-functions being zero, the critical line has 
\begin{align}
\alpha_{u_{2}} & ~=~\frac{3}{4(1+x_{F})}\left(\pm\sqrt{1+4x_{F}}-1\right)\alpha_{g}~~\longrightarrow\begin{cases}
\mbox{$\frac{3(-1\pm\sqrt{23})}{26}$}\alpha_{g} & :\,x_{F}\rightarrow22/4\\
\mbox{$\frac{3(-1\pm\sqrt{22})}{25}$}\alpha_{g} & :\,x_{F}\rightarrow21/4
\end{cases}~,\nonumber \\
\alpha_{w_{2}} & ~=~\frac{1}{32(6-x_{F})}\left(3\pm\sqrt{6x_{F}-27}\right)\alpha_{g}~~\longrightarrow\begin{cases}
\mbox{$\frac{3\pm\sqrt{6}}{16}\alpha_{g}$} & \,:\,x_{F}\rightarrow22/4\\
\mbox{$\frac{2\pm\sqrt{2}}{16}\alpha_{g}$} & \,:\,x_{F}\rightarrow21/4
\end{cases}\,\,~.
\end{align}
The results of \cite{Abel:2017ujy} are recovered 
when $x=22/4$ as expected, and evidently the behaviour is not qualitatively altered by the presence
of order $N$ scalars. There are two beta-functions remaining:
\begin{align}
\beta_{u_{1}} &~ =~16\alpha_{u_{1}}^{2}+16\alpha_{u_{2}}\alpha_{u_{1}}+3\alpha_{u_{2}}^{2}+2\alpha_{u_{1}}\alpha_{y}\,~,\nonumber \\
\beta_{w_{1}} & ~=~16x_{S}\alpha_{w_{1}}^{2}+16(2+x_{S})\alpha_{w_{1}}\alpha_{w_{2}}+24\alpha_{w_{2}}^{2}-6\alpha_{g}\alpha_{w_{1}}+\mbox{$\frac{3}{16}$}\alpha_{g}^{2}\,~.
\end{align}
Solving for these beta functions being zero, we find that there are only real solutions
for $\alpha_{w_{1}}$ and $\alpha_{u_{1}}$ for the branches,  
\begin{align}
\alpha_{w_{2}} & 
%=\,\mbox{$\frac{3-\sqrt{6x_{F}-27}}{32(6-x_{F})}$}\alpha_{g}\nonumber \\
  ~=~\mbox{$\frac{2-\sqrt{2}}{16}$}\alpha_{g}~, \nonumber \\
\alpha_{u_{2}} & 
%=\,\mbox{$ \frac{3(\sqrt{1+4x_{F}}-1)}{4(1+x_{F})}$} \alpha_{g}\nonumber \\
  ~=~\mbox{$\frac{3 (\sqrt{22}-1)}{25}$}\alpha_{g}\,\,,
\end{align}
(or the equivalent for $x_{F}=22/4$). Inserting the appropriate
$\alpha_{w_{2}}$ solution we find a fixed point at
\begin{align}
\alpha_{u_{1}} & ~~=~~\begin{cases}
\frac{-6\sqrt{23}\pm3\sqrt{20+6\sqrt{23}}}{104}\alpha_{g} & \,:\,x_{F}\rightarrow22/4\\
\frac{-6\sqrt{22}\pm3\sqrt{19+6\sqrt{22}}}{100}\alpha_{g} & \,:\,x_{F}\rightarrow21/4
\end{cases}\,\,,\nonumber \\
\alpha_{w_{1}} & ~~=~~\begin{cases}
\frac{17\sqrt{6}-36}{128}\alpha_{g} & \,:\,x_{F}\rightarrow22/4\\
\frac{3\pm\sqrt{3(4\sqrt{2}-5)})}{16\sqrt{2}}\alpha_{g} & \,:\,x_{F}\rightarrow21/4
\end{cases}\,\,.
\end{align}
Note that for $x_{F}=22/4$ there is only one solution for $\alpha_{w_{1}}$
because the quadratic term in $\beta_{w_{1}}$ is proportional to
$x_{S}$. 

The appropriate fixed point is then determined by stability. The effective quadratic couplings are  
\begin{equation}
\lambda_{h_0} ~=~ \frac{96 \pi^2 }{N_F^2}\left(\alpha_{u_1}+\alpha_{u_2}\right) ~~;~~~ \lambda_{\tilde{Q}} ~=~  \frac{96 \pi^2 }{N^2}\
\left(\alpha_{w_1}+\frac{N}{N_S} \alpha_{w_2} \right) ~~,
\end{equation}
where the corresponding quartic term in the potential of the canonically normalized diagonal component of the field, $\phi$,  is $\sim \lambda \phi^2/4!$~.  
The quadratic term for $\tilde{Q}$ is positive regardless of which $w_1$ branch is chosen, but imposing  $\lambda_{h_0}>0 $ selects the  
$-6\sqrt{22}+3\sqrt{19-6\sqrt{22}}$ ~branch
for $\alpha_{u_{1}}$: it has negative $\alpha_{u_{1}}=-0.08\alpha_g $, but the overall quadratic term is still positive. 
There are then two possible fixed points one for 
each branch of $w_1$.
Let us summarise by collecting the values for the pair of consistent stable 
fixed points; \\

\noindent \underline{ \it $N_{S}=N-3$ implies $N_{S}/N\rightarrow1$ in the Veneziano
limit ($x_{F}\rightarrow21/4$):}

\begin{align}
\alpha_{u_{2}}\,\, & =\,\,\mbox{$\frac{3(\sqrt{22}-1)}{25}$}\alpha_{g}\nonumber \protect\\
\alpha_{w_{2}}\,\, & =\,\,\mbox{$\frac{2-\sqrt{2}}{16}$}\alpha_{g}\nonumber \protect\\
\alpha_{u_{1}}\,\, & =\,\,\mbox{$\frac{-6\sqrt{22}+3\sqrt{19+6\sqrt{22}}}{100}$}\alpha_{g}\nonumber \protect\\
\alpha_{w_{1}}\,\, & =\,\,\mbox{$\frac{3\pm\sqrt{3(4\sqrt{2}-5)})}{16\sqrt{2}}$}\alpha_{g} ~~.
\end{align}\\

\noindent \underline{ \it $N_{S}=const$ implies $N_{S}/N\rightarrow1$ in the Veneziano
limit ($x_{F}\rightarrow22/4$):}

\begin{align}
\alpha_{u_{2}}\,\, & =\,\,\mbox{$\frac{3(\sqrt{23}-1)}{26}$}\alpha_{g}\nonumber \protect\\
\alpha_{w_{2}}\,\, & =\,\,\mbox{$\frac{3-\sqrt{6}}{16}$}\alpha_{g}\nonumber \protect\\
\alpha_{u_{1}}\,\, & =\,\,\mbox{$\frac{-6\sqrt{23}+3\sqrt{20+6\sqrt{23}}}{104}$}\alpha_{g}\nonumber \protect\\
\alpha_{w_{1}}\,\, & =\,\,\mbox{$\frac{17\sqrt{6}-36}{128}$}\alpha_{g}~~.
\end{align}
Finally one can solve for the coupling $\alpha_g(t)$ itself. The solution is most easily defined in terms of 
\begin{equation}
\omega(t)~=~\frac{\alpha^*_g}{\alpha_g(t)}-1~~ .
\end{equation}
One finds
\begin{equation}
\omega(t)~=~W\left[ \omega(0)e^{\omega(0)}\,e^{-\frac{4}{3}\alpha_g^*\epsilon t} \right]\, \, ,
\end{equation}   
where the Lambert $W$ function is given by $z=W[ze^z]$ (as is evident from setting $t=0$). \\

%%%%%%%%%%%%% 

%XXXXXX

\begin{centering}

%NEW BELOW 
\begin{figure}[h]
\begin{fmffile}{bubbles}
    \begin{fmfgraph}(90,90)
       \fmfleft{i,i1,i2}
       \fmfright{o,o1,o2}
        \fmf{phantom}{i1,v1,v2,o1}
        \fmf{photon}{i1,v1}
        \fmf{photon}{o1,v2}        
        \fmf{phantom}{i,v3,o}
        \fmf{phantom}{i2,v4,o2}
        \fmf{plain,right,tension=0.01}{v1,v2,v1}
     \end{fmfgraph}
%%%%%%%%
     \hspace{-2.0cm} 
     $\tilde{\alpha} $
     \hspace{2.3cm}
%%%%%%%%
    \begin{fmfgraph}(90,90)
       \fmfleft{i,i1,i2}
       \fmfright{o,o1,o2}
        \fmf{phantom}{i1,v1,v2,o1}
        \fmf{photon}{i1,v1}
        \fmf{photon}{o1,v2}        
        \fmf{phantom}{i,v3,o}
        \fmf{phantom}{i2,v4,o2}
                \fmf{photon}{v3,v4}
        \fmf{plain,right,tension=0.01}{v1,v2,v1}
     \end{fmfgraph}
%%%%%%%%
     \hspace{-2.5cm} 
     $\frac{1}{N_f}\tilde{\alpha}^{2 }  $
     \hspace{1.2cm}
%%%%%%%%
    \begin{fmfgraph}(90,90)
       \fmfleft{i,i1,i2}
       \fmfright{o,o1,o2}
        \fmf{phantom}{i1,v1,v2,o1}
        \fmf{photon}{i1,v1}
        \fmf{photon}{o1,v2}        
        \fmf{phantom,tension=0.67}{i,v3,o}
        \fmf{phantom,tension=0.67}{i2,v4,o2}
                \fmf{phantom}{v3,v5,v6,v4}
                \fmf{photon}{v3,v5}
                \fmf{photon}{v6,v4}
        \fmf{plain,right,tension=0.01}{v1,v2,v1}
        \fmf{plain,right,tension=0.5}{v5,v6,v5}
%%%%%%%%
     \hspace{-2.2cm} 
     $\frac{1}{N_f}\tilde{\alpha}^{3}  $
     \hspace{0.7cm}
%%%%%%%%
     \end{fmfgraph}
    \begin{fmfgraph}(10,90)
       \fmfleft{i}
       \fmfright{o}
                \fmf{dots}{i,o}
     \end{fmfgraph}
    \begin{fmfgraph}(90,90)
       \fmfleft{i,i1,i2}
       \fmfright{o,o1,o2}
        \fmf{phantom}{i1,v1,v2,o1}
        \fmf{photon}{i1,v1}
        \fmf{photon}{o1,v2}        
        \fmf{phantom,tension=0.42}{i,v3,o}
        \fmf{phantom,tension=0.42}{i2,v4,o2}
                \fmf{phantom}{v3,v5,v6,v7,v8,v4}
                %\fmf{phantom}{v5,v6}
                %\fmf{phantom}{v7,v8}
                 \fmf{photon}{v3,v5}
                \fmf{photon}{v8,v4}
                \fmf{dots,tension=0.5}{v6,v7}
                 \fmf{plain,right,tension=0.01}{v1,v2,v1}
        \fmf{plain,right,tension=1}{v5,v6,v5}
        \fmf{plain,right,tension=1}{v7,v8,v7}
%%%%%%%%
     \hspace{-2.5cm} 
     $\frac{1}{N_f}\tilde{\alpha}^{(L-1)} $
     \hspace{1.3cm}
%%%%%%%%
     \end{fmfgraph}

%%%%%%%%
     \hspace{-1.cm} 
    \mbox{ }
 %%%%%%%%
    \begin{fmfgraph}(90,90)
     \end{fmfgraph}
    \begin{fmfgraph}(90,90)
       \fmfleft{i,i1,i2}
       \fmfright{o,o1,o2}
        \fmf{phantom}{i1,v1,v2,o1}
        \fmf{photon}{i1,v1}
        \fmf{photon}{o1,v2}        
        \fmf{phantom,tension=0.3}{i,v3,v4,o}
        \fmf{phantom,tension=0.7}{i1,v3,v4,o1}
        \fmf{phantom,tension=0.8}{i1,v3}
        \fmf{phantom,tension=0.8}{v4,o1}
        \fmf{plain,right,tension=0.01}{v1,v2,v1}
                \fmf{photon,tension=0.02}{v3,v4}
%%%%%%%%
     \hspace{-2.2cm} 
     $\frac{1}{N_f}\tilde{\alpha}^{2} $
     \hspace{0.9cm}
%%%%%%%%
     \end{fmfgraph}
    \begin{fmfgraph}(90,90)
       \fmfleft{i,i1,i2}
       \fmfright{o,o1,o2}
        \fmf{phantom}{i1,v1,v2,o1}
        \fmf{photon}{i1,v1}
        \fmf{photon}{o1,v2}        
        \fmf{phantom,tension=0.1}{i,v3,v5,v6,v4,o}
        \fmf{phantom,tension=0.3}{i1,v3,v5,v6,v4,o1}
        \fmf{phantom,tension=0.2}{i1,v3}
        \fmf{phantom,tension=0.2}{v4,o1}
        \fmf{plain,right,tension=0.01}{v1,v2,v1}
                \fmf{photon,tension=0.64}{v3,v5}
                \fmf{photon,tension=0.64}{v6,v4}
                \fmf{plain,right,tension=0.25}{v5,v6,v5}
%%%%%%%%
     \hspace{-2.1cm} 
     $\frac{1}{N_f}\tilde{\alpha}^{\, 3} $
     \hspace{0.6cm}
%%%%%%%%
     \end{fmfgraph}
    \begin{fmfgraph}(10,90)
       \fmfleft{i}
       \fmfright{o}
                \fmf{dots}{i,o}
     \end{fmfgraph}  
    \begin{fmfgraph}(90,90)
       \fmfleft{i,i1,i2}
       \fmfright{o,o1,o2}
        \fmf{phantom}{i1,v1,v2,o1}
        \fmf{photon}{i1,v1}
        \fmf{photon}{o1,v2}        
        \fmf{phantom,tension=0.04}{i,v3,v5,v7,v8,v6,v4,o}
        \fmf{phantom,tension=0.13}{i1,v3,v5,v7,v8,v6,v4,o1}
        \fmf{phantom,tension=0.1}{i1,v3}
        \fmf{phantom,tension=0.1}{v4,o1}
        \fmf{plain,right,tension=0.01}{v1,v2,v1}
                \fmf{photon,tension=0.5}{v3,v5}
                \fmf{photon,tension=0.5}{v6,v4}
                \fmf{plain,right,tension=0.5}{v5,v7,v5}
                \fmf{plain,right,tension=0.5}{v6,v8,v6}
                \fmf{dots,tension=0.3}{v7,v8}
%%%%%%%%
     \hspace{-2.0cm} 
     $\frac{1}{N_f}\tilde{\alpha}^{(L-1)} $
     \hspace{-0.5cm}
%%%%%%%%
     \end{fmfgraph}
\end{fmffile}
\caption{\label{bubbles1} \it One-loop diagram and the leading resummed pole contributions for the $SU(2)_L\times SU(2)_R$ fixed points, where $\tilde{\alpha}= N_f\,g^2_{SU(2)}/16\pi^2$. {\it The plain lines represent both quarks and scalars}.}
\end{figure}
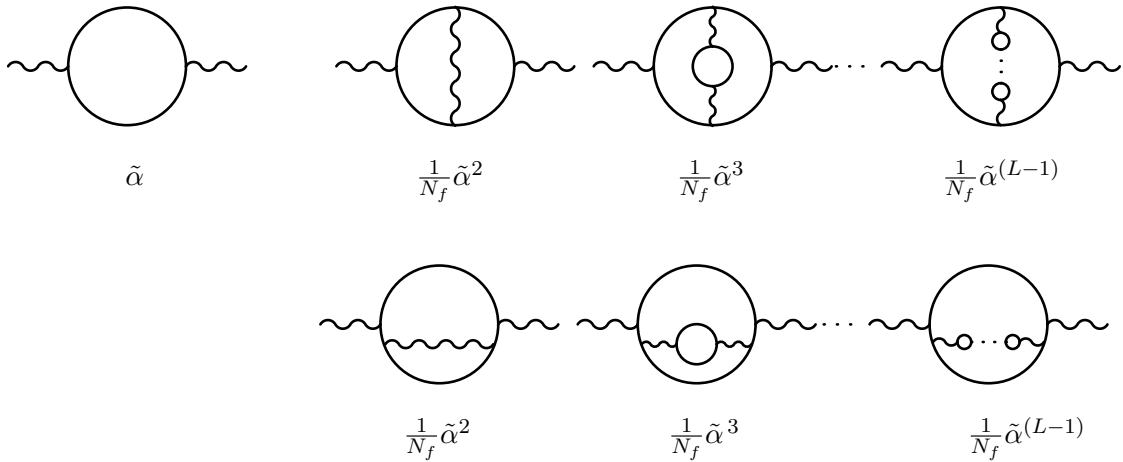

\begin{figure}[h!]
\begin{fmffile}{bubbles2}
         \begin{fmfgraph}(90,90)
       \fmfleft{i,i1,i2}
       \fmfright{o,o1,o2}
        \fmf{phantom}{i1,v1,v2,o1}
        \fmf{photon}{i1,v1}
        \fmf{photon}{o1,v2}        
        \fmf{phantom,tension=0.04}{i,v3,v5,v7,v11,v8,v6,v4,o}
        \fmf{phantom,tension=0.13}{i1,v3,v5,v7,v8,v6,v4,o1}
        \fmf{phantom,tension=0.09}{i1,v3}
        \fmf{phantom,tension=0.09}{v4,o1}
        \fmf{plain,right,tension=0.002}{v1,v2,v1}
                \fmf{photon,tension=0.5}{v3,v5}
                \fmf{photon,tension=0.5}{v6,v4}
                \fmf{plain,right,tension=0.5}{v5,v7,v5}
                \fmf{plain,right,tension=0.5}{v6,v8,v6}
                \fmf{dots,tension=0.2}{v7,v8}
         \fmffreeze       
         \fmf{phantom,tension=0.01}{i2,v9,o2}
         \fmf{photon,tension=0.017}{v9,v11}
              \end{fmfgraph}
%%%%%%%%
     \hspace{-2.6cm} 
     $\frac{1}{N^2_f}\tilde{\alpha}^{(L-1)} $
     \hspace{-0.cm}
     
%%%%%%%%
    \begin{fmfgraph}(90,90)
       \fmfleft{i,i1,i2}
       \fmfright{o,o1,o2}
        \fmf{phantom}{i1,v1,v2,o1}
        \fmf{photon}{i1,v1}
        \fmf{photon}{o1,v2}        
        \fmf{phantom}{i,v3,o}
        \fmf{phantom}{i2,v4,o2}
                \fmf{gluon}{v4,v3}
        \fmf{plain,right,tension=0.01}{v1,v2,v1}
     \end{fmfgraph}
     %%%%%%%%
     \hspace{-2.8cm} 
     $\alpha_g \tilde{\alpha}\sim \epsilon\tilde{\alpha} $
     \hspace{0.8cm}
%%%%%%%%
    \begin{fmfgraph}(90,90)
       \fmfleft{i,i1,i2}
       \fmfright{o,o1,o2}
        \fmf{phantom}{i1,v1,v2,o1}
        \fmf{photon}{i1,v1}
        \fmf{photon}{o1,v2}        
        \fmf{phantom}{i,v3,o}
        \fmf{phantom}{i2,v4,o2}
                \fmf{plain}{v4,v3}
        \fmf{plain,right,tension=0.01}{v1,v2,v1}
     \end{fmfgraph}
     %%%%%%%%
     \hspace{-2.8cm} 
     $\alpha_y \tilde{\alpha}\sim \epsilon\tilde{\alpha} $
     \hspace{1.0cm}
%%%%%%%%
   \begin{fmfgraph}(10,90)
       \fmfleft{i}
       \fmfright{o}
                \fmf{dots}{i,o}
     \end{fmfgraph}  
        \begin{fmfgraph}(90,90)
       \fmfleft{i,i1,i2}
       \fmfright{o,o1,o2}
        \fmf{phantom}{i1,v1,v2,o1}
        \fmf{photon}{i1,v1}
        \fmf{photon}{o1,v2}        
        \fmf{phantom,tension=0.04}{i,v3,v5,v7,v11,v8,v6,v4,o}
        \fmf{phantom,tension=0.13}{i1,v3,v5,v7,v8,v6,v4,o1}
        \fmf{phantom,tension=0.09}{i1,v3}
        \fmf{phantom,tension=0.09}{v4,o1}
        \fmf{plain,right,tension=0.001}{v1,v2,v1}
                \fmf{photon,tension=0.5}{v3,v5}
                \fmf{photon,tension=0.5}{v6,v4}
                \fmf{plain,right,tension=0.5}{v5,v7,v5}
                \fmf{plain,right,tension=0.5}{v6,v8,v6}
                \fmf{dots,tension=0.17}{v7,v8}
         \fmffreeze       
         \fmf{phantom,tension=0.01}{i2,v9,o2}
         \fmf{gluon,tension=0.016}{v9,v11}
    \end{fmfgraph}
 %%%%%%%%
     \hspace{-2.7cm} 
     $\frac{\epsilon}{N_f}\tilde{\alpha}^{(L-1)} $
     \hspace{0.8cm}
%%%%%%%%
       \begin{fmfgraph}(90,90)
       \fmfleft{i,i1,i2}
       \fmfright{o,o1,o2}
        \fmf{phantom}{i1,v1,v2,o1}
        \fmf{photon}{i1,v1}
        \fmf{photon}{o1,v2}        
        \fmf{phantom,tension=0.04}{i,v3,v5,v7,v11,v8,v6,v4,o}
        \fmf{phantom,tension=0.13}{i1,v3,v5,v7,v8,v6,v4,o1}
        \fmf{phantom,tension=0.09}{i1,v3}
        \fmf{phantom,tension=0.09}{v4,o1}
        \fmf{plain,right,tension=0.001}{v1,v2,v1}
                \fmf{photon,tension=0.5}{v3,v5}
                \fmf{photon,tension=0.5}{v6,v4}
                \fmf{plain,right,tension=0.5}{v5,v7,v5}
                \fmf{plain,right,tension=0.5}{v6,v8,v6}
                \fmf{dots,tension=0.17}{v7,v8}
         \fmffreeze       
         \fmf{phantom,tension=0.01}{i2,v9,o2}
         \fmf{plain,tension=0.016}{v9,v11}
%%%%%%%%
     \hspace{-2.5cm} 
     $\frac{\epsilon}{N_f}\tilde{\alpha}^{(L-1)} $
     \hspace{-0.cm}
%%%%%%%%
     \end{fmfgraph}
     
%%%%%%%%
    \begin{fmfgraph}(90,90)
       \fmfleft{i,i1,i2}
       \fmfright{o,o1,o2}
        \fmf{phantom}{i1,v1,v2,o1}
        \fmf{photon}{i1,v1}
        \fmf{photon}{o1,v2}        
        \fmf{phantom}{i,v3,o}
        \fmf{phantom}{i2,v4,o2}
        \fmf{plain,right=0.5,tension=0.5}{v3,v4}
        \fmf{plain,left=0.5,tension=0.5}{v3,v4}
        \fmf{plain,right,tension=0.01}{v1,v2,v1}
     \end{fmfgraph}
     %%%%%%%%
     \hspace{-3cm} 
     $\alpha_{u_2}^2\tilde{\alpha}\sim \epsilon^2\tilde{\alpha} $
     \hspace{1.cm}
%%%%%%%%
   \begin{fmfgraph}(10,90)
       \fmfleft{i}
       \fmfright{o}
                \fmf{dots}{i,o}
     \end{fmfgraph}  
        \begin{fmfgraph}(90,90)
       \fmfleft{i,i1,i2}
       \fmfright{o,o1,o2}
        \fmf{phantom}{i1,v1,v2,o1}
        \fmf{photon}{i1,v1}
        \fmf{photon}{o1,v2}        
        \fmf{phantom,tension=0.04}{i,v3,v5,v7,v11,v8,v6,v4,o}
        \fmf{phantom,tension=0.13}{i1,v3,v5,v7,v8,v6,v4,o1}
        \fmf{phantom,tension=0.09}{i1,v3}
        \fmf{phantom,tension=0.09}{v4,o1}
        \fmf{plain,right,tension=0.001}{v1,v2,v1}
                \fmf{photon,tension=0.5}{v3,v5}
                \fmf{photon,tension=0.5}{v6,v4}
                \fmf{plain,right,tension=0.5}{v5,v7,v5}
                \fmf{plain,right,tension=0.5}{v6,v8,v6}
                \fmf{dots,tension=0.17}{v7,v8}
         \fmffreeze       
         \fmf{phantom,tension=0.01}{i2,v9,o2}
        \fmf{plain,right=0.5,tension=0.008}{v9,v11}
        \fmf{plain,left=0.5,tension=0.008}{v9,v11}
%%%%%%%%
     \hspace{-2.5cm} 
     $\frac{\epsilon^2}{N_f}\tilde{\alpha}^{(L-1)} $
     \hspace{-0.cm}
%%%%%%%%
     \end{fmfgraph}
\end{fmffile}
\caption{\label{bubbles2} \it Sub-leading bubbles in the renormalisation of the $SU(2)_{L/R}$ gauge couplings 
 ({\it where plain lines represent quarks and/or scalars})  are suppressed 
with respect to the terms in the resummation in figure \ref{bubbles1}. 
The first diagrams exist also in the pure $SU(2)_{L/R}$ theory and are suppressed by a factor of $1/N_f$: this admits 
the procedure of  \cite{Holdom} that establishes a fixed point by balancing the resummed pole 
of figure \ref{bubbles1} against the one-loop diagram. On the second row, the insertion of an $SU(N)$ ``gluon" line on the quark loops, or a Higgs scalar via Yukawa couplings,  gives a factor $\frac{g^2}{(4\pi)^2} \sum_A (T^A T^A) \sim \frac{g^2N}{(4\pi)^2}  \sim \epsilon $ or a factor $\frac{y^2N_F}{(4\pi)^2}  \sim \epsilon $ respectively, compared to diagrams with the same power of $\tilde{\alpha}$ 
in  figure \ref{bubbles1}, where the $\epsilon $ scalings apply when one is near the fixed points of $\alpha_g,\alpha_y,\alpha_{u_1},\alpha_{u_2}$ couplings. Note that due to the large number of $SU(2)$ doublets, $\epsilon$ is of order $5/N_f$. On the third row, the introduction of a pair of scalars with quartic interactions introduced terms suppressed even more,  by factors of order $\alpha_{u_2}^2\sim \epsilon^2$.}
\end{figure}
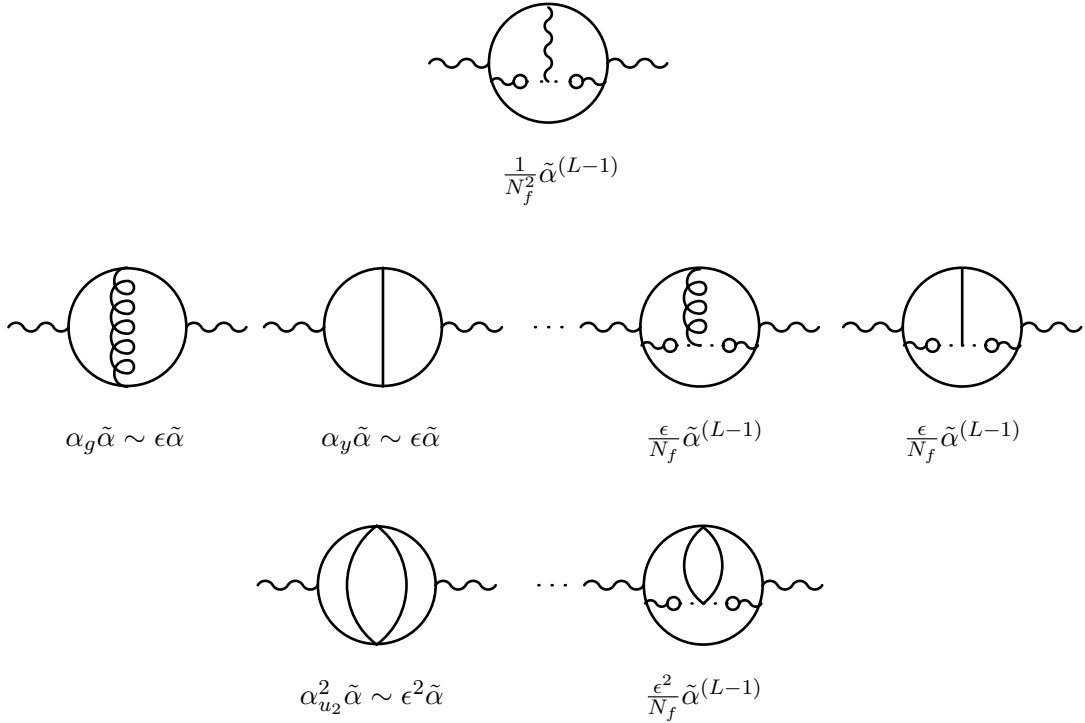
\end{centering}

\begin{centering}

%NEW BELOW 
\begin{figure}[h!]
\begin{fmffile}{bubbles3}
    \begin{fmfgraph}(90,90)
       \fmfleft{i,i1,i2}
       \fmfright{o,o1,o2}
        \fmf{phantom}{i1,v1,v2,o1}
        \fmf{gluon}{i1,v1}
        \fmf{gluon}{o1,v2}        
        \fmf{phantom}{i,v3,o}
        \fmf{phantom}{i2,v4,o2}
        \fmf{plain,right,tension=0.01}{v1,v2,v1}
     \end{fmfgraph}
%%%%%%%%
     \hspace{-2.3cm} 
     $x_F\alpha_g $
     \hspace{1.5cm}
%%%%%%%%
    \begin{fmfgraph}(90,90)
       \fmfleft{i,i1,i2}
       \fmfright{o,o1,o2}
        \fmf{phantom}{i1,v1,v2,o1}
        \fmf{gluon}{i1,v1}
        \fmf{gluon}{o1,v2}        
        \fmf{phantom}{i,v3,o}
        \fmf{phantom}{i2,v4,o2}
                \fmf{gluon}{v3,v4}
        \fmf{plain,right,tension=0.01}{v1,v2,v1}
     \end{fmfgraph}
%%%%%%%%
     \hspace{-2.7cm} 
     $x_F \,\alpha^2_g \sim \epsilon^2 $
     \hspace{.7cm}
%%%%%%%%
%%%%%%%%
    \begin{fmfgraph}(90,90)
       \fmfleft{i,i1,i2}
       \fmfright{o,o1,o2}
        \fmf{phantom}{i1,v1,v2,o1}
        \fmf{gluon}{i1,v1}
        \fmf{gluon}{o1,v2}        
        \fmf{phantom}{i,v3,o}
        \fmf{phantom}{i2,v4,o2}
                \fmf{plain}{v3,v4}
        \fmf{plain,right,tension=0.01}{v1,v2,v1}
     \end{fmfgraph}
%%%%%%%%
     \hspace{-3.0cm} 
     $x_F\, \alpha_{y}\alpha_g \sim \epsilon^2  $
     \hspace{.6cm}
%%%%%%%%
%%%%%%%%
    \begin{fmfgraph}(90,90)
       \fmfleft{i,i1,i2}
       \fmfright{o,o1,o2}
        \fmf{phantom}{i1,v1,v2,o1}
        \fmf{gluon}{i1,v1}
        \fmf{gluon}{o1,v2}        
        \fmf{phantom}{i,v3,o}
        \fmf{phantom}{i2,v4,o2}
                \fmf{photon}{v3,v4}
        \fmf{plain,right,tension=0.01}{v1,v2,v1}
     \end{fmfgraph}
%%%%%%%%
     \hspace{-3.7cm} 
     $\frac{1}{N_fN}\tilde{\alpha}\alpha_g \,\sim\, \frac{\epsilon}{N^2_f}\tilde{\alpha} \,\lesssim\, \epsilon^3 \frac{\tilde{\alpha}}{25} $
     \hspace{0.2cm}
%%%%%%%%
         \begin{fmfgraph}(10,90)
       \fmfleft{i}
       \fmfright{o}
                \fmf{dots}{i,o}
     \end{fmfgraph}  
\end{fmffile}
\caption{\label{bubbles3} \it Factors contributing to the beta functions of the $SU(N)$ gauge coupling to two-loops (which get multiplied by an overall $\alpha_g$ factor in $\beta_{g}$). The leading term is of course cancelled to order $\epsilon^2$ against the gauge loops in the Banks-Zaks limit by the choice of colours and flavours. Noting that $\epsilon \gtrsim 1/N \gtrsim 5/N_f$, the $SU(2)_{L/R}$ gauging can be neglected in the $N_f\rightarrow \infty$ limit. {\it The plain lines represent both quarks and scalars}.}
\end{figure}
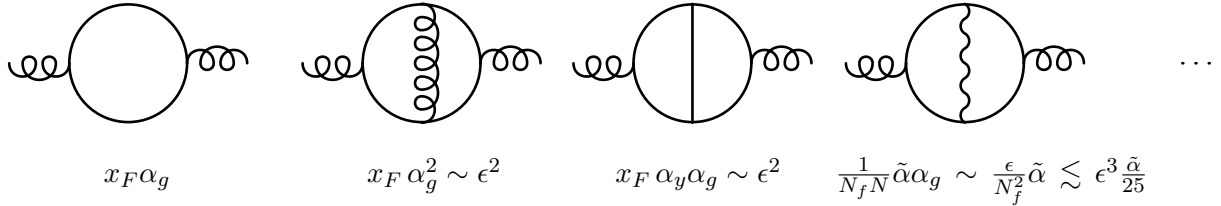
\end{centering}

\section{The electroweak $SU(2)_L\times SU(2)_R$ UV fixed point, $A''$,  and its decoupling in the Banks-Zaks limit}

We now turn to the gauging of the electroweak couplings $SU(2)_L\times SU(2)_R$. Consistency of the  picture, namely that there is an overall UV fixed point,  requires that these couplings join in with the fixed point behaviour in the UV. In this section we discuss the existence of fixed points for these factors which 
are effectively $SU(2)$ gauge theories in  a large ``flavour''  expansion (with order $N$ flavours of electroweak fundamental). We then establish that, crucially in the 
Banks-Zaks limit we are considering, the flow to the $SU(N)$ fixed point, $A'$, and to the $SU(2)_L\times SU(2)_R$ fixed point, $A''$,  can be  established independently of one another. In other words, a large colour and large flavour Banks-Zaks  UV fixed point can coexist and not-interfere with a large flavour fixed point of the kind established in \cite{Holdom,Pica:2010xq,Litim:2014uca}. In terms of figure \ref{fig:flow}, this means that the profile of the trajectory in the $g,y$ plane is independent of $g'$, while the flow in the  $g'$ direction as a function of $t$ is independent of $g,y$. (Note that a complimentary approach would be to add additional coloured multiplets to 
  achieve ``large flavour'' fixed points for {\it all} the gauge groups \cite{Mann:2017wzh}.)

First let us address the existence of UV fixed points for the $SU(2)$ gauge couplings in the presence of a large effective number of $SU(2)$ flavours, $N_f$. By ``effective'' we mean that, as we shall see, in the leading diagrams, scalar and fermion bubbles contribute equivalently up to a factor, so their contributions always appear in the same linear combination.  
 (Note that $N_f$ for these fixed points is not to be confused with the previous $N_F$.)  
 A $1/N_f$  expansion can be organised in terms of 
 \begin{equation}
 \tilde{\alpha} ~=~ \frac{N_f g^{\prime 2}
}{(4\pi)^2}\,\,\, ~,
\end{equation} 
with $g'$ standing for $g_{SU(2)_L}$ or $g_{SU(2)_R}$. Typical terms contributing to the beta-function are shown in figure \ref{bubbles1}. 
They  can be resummed (see for example the review in \cite{Holdom}), and one finds \cite{Pica:2010xq}
\begin{equation}
\label{bebe}
\frac{3}{4}\frac{\beta_{\tilde{\alpha}}}{{\tilde{\alpha}}^2}= 1+\frac{H(\tilde{\alpha})}{N_f}+{\cal O}(N_f^{-2}) \,\, ,
\end{equation}
where the additional terms, suppressed by at least a factor of $1/N_f^2 $, arise from diagrams such as the 
class shown on the first line of figure \ref{bubbles2}.
From  figure \ref{bubbles1} it is clear that contributions from scalar and quark loops are simply additive in the resummation, with the number of $SU(2)_{L/R}$ quark doublets being $3 N/2$, and scalar doublets being  $3 N_F/4 $ for $SU(2)_L$ and $3 N_F/4 + N/4$ for $SU(2)_R$, in the SM embedding we are considering. Therefore setting $N_F \approx 21/4$,
  $SU(2)_L$ has $N_f\approx 87N/16$ while  $SU(2)_R$ has $N_f\approx 91N/16$. 
  
 The important point about the function  $H(\tilde{\alpha})$ is that it has a {\it negative} logarithmic singularity at \begin{equation}
\tilde{\alpha}_0~=~ \frac{3}{2}\,\,\, .
\end{equation}
Thus one can always find a solution to $\beta_{\tilde{\alpha}}(\tilde{\alpha})=0$ at $\tilde{\alpha}_*$ somewhat below this value. For values of $N_f\gtrsim 5$, $\tilde{\alpha}$ runs in the UV rapidly to a value $\tilde{\alpha}_*$ that is in fact exponentially close to $3/2$ \cite{Pica:2010xq,Litim:2014uca}. Indeed the form of the singularity 
is \begin{equation}
H(\tilde{\alpha})  ~=~ \frac{1}{4}\log |3-2\tilde{\alpha} |+ \mbox{constant ,} 
\end{equation}
so  one has $\tilde{\alpha}_*  ~ = ~\frac{3}{2} - C e^{-4N_f} $,
for some constant $C$. For even modest $N_f$ (note that for $N=10$ one has $N_f\sim 50$) the exponential term is completely negligible.

The existence of such large $N_f$ fixed points is  well established, modulo the somewhat trivial additional contribution of scalars in the loops. However let us now consider the effect of the couplings that are being turned on in the rest of the theory.
In the presence of the $SU(N)$ gauging, the Yukawa  couplings, and the scalar quartic terms, there is the possibility of disturbing the eletroweak fixed point. However by following the power-counting outlined in figures \ref{bubbles1} and \ref{bubbles2}, one finds that such contributions are suppressed by order $\epsilon $ with respect to the terms in the resummation with the same powers of $\tilde{\alpha}$. Thus as long as the terms proportional to $\epsilon$ do not themselves have a pole at smaller values of $\tilde{\alpha}$, this induces a completely negligible shift (in the exponential term) in the value of  $\tilde{\alpha}_*$. That is, near the pole the beta function is shifted as 
\begin{equation}
\label{bebe2}
\frac{3}{4}\frac{\beta_{\tilde{\alpha}}}{{\tilde{\alpha}}^2}= 1+c(\tilde{\alpha})  \, \epsilon+ \frac{H(\tilde{\alpha})}{N_f}+\ldots \,\, ,
\end{equation}
for some $c(\tilde{\alpha}_0) $ of order unity. Solving for $\beta(\tilde{\alpha})=0$ one finds that $\tilde{\alpha}_*  ~ = ~\frac{3}{2} - C e^{-4N_f(1+c\epsilon)} $, in other words the fixed point is barely shifted, because the gauged colour coupling adds a subleading contribution to the beta function in the Veneziano limit. 

In this limit therefore the $\alpha_{{SU(2)}_{L/R}}$ couplings have the same UV fixed points as the theory with an effective flavour number $N_f$, without the $SU(N)$ gauging, Yukawas and scalar quartics. Of course the running away from the fixed point will be altered by the presence of the $\epsilon$ term, but here the running is dominated by the leading one-loop term of \eqref{bebe2}, and thus one  expects the $SU(N)$ gauging to cause a (two-loop) shift in the RG trajectory of $\tilde{\alpha} $ that is suppressed by a factor $\tilde{\alpha}\epsilon$. 

Remarkably the electroweak $\alpha_{{SU(2)}_{L/R}}$ gauging also decouples from the UV fixed points of the gauge Yukawa sector except leaving a residual possible shift in the value of the fixed point. This is easier to treat  because firstly the fixed point can be determined by the leading diagrams, and secondly 
only a finite number (i.e. 6) of flavours are gauged under the electroweak symmetry. 
As an example we display the leading contributions to the beta function of $\alpha_g$ in figure \ref{bubbles3}.
 The last diagram of this figure shows a new contribution from the internal insertion of a flavour gauge boson. Inserting parameters and 
 noting that $\epsilon \gtrsim 1/N $ (with from our earlier discussion $N_f \sim 5N$), with $\tilde{\alpha}\approx 3/2$, 
  these new diagrams are suppressed by a factor of order $\epsilon/25$ with respect to the other two-loop contribution in the 
  beta function of $\alpha_g$. They result in a small shift in the fixed point value that is comparable to that coming from the 
  three-loop diagrams that are already being neglecting.  
  
  \section{Discussion: towards realistic phenomenology}

  The dislocation of the electroweak and strong fixed points in the Veneziano limit is an attractive feature of the present set-up. 
  It implies that there is always a large enough number of colours and flavour for which the fixed point is guaranteed to exist. It is interesting nevertheless to insert finite values to ascertain the phenomenological viability of the set-up, in particular, whether it allows a consistent set of coupling values for reasonable values of colours and flavours. We shall take $N=10$ as our representative example. 
  Before continuing we should warn the reader that we are not looking for an exact reproduction of the SM values, because we have 
  not yet broken flavour degeneracy. There are still 3 Higgs pairs that will be driven to acquire a degenerate VEV by a negative $m_0^2$, and as there is still unbroken flavour symmetry one  expects to find Goldstone scalars 
  in the spectrum as well.  Ultimately one would like to break the remaining flavour symmetry to induce hierarchies in the fermion masses, and to leave a single Higgs dominant in the electroweak symmetry breaking. The purpose of the present discussion is  to demonstrate that the 
  prospects for a full SM phenomenology are encouraging, with  these more detailed aspects being left to future work.
    
First we should note that in the Veneziano limit both the electroweak and the $SU(N)$ coupling are weakly coupled in the UV. 
By \eqref{eq:astar} the fixed point value of the strong gauge coupling is  $\alpha^*_g = 25/18 \epsilon$. An $\epsilon $ of order $0.075$ (well inside the domain of attraction of the fixed point \cite{Litim:2014uca}) can be achieved with $N=10$, by taking $N_S=N-3=7$ and $N_F=54$. For this example the actual $SU(N)$ coupling at the fixed point is then  \[\alpha^*_s ~=~ 4\pi \alpha_g^* /N ~=~ 4\pi \frac{25\epsilon}{18N} ~=~ 0.13~.\] Note that the renormalisation of this coupling is  weak, so a remarkably consistent value is achievable even with rather large numbers of colours. For the quartic couplings,  taking a Higgs mass of $125$~GeV and a VEV $h_0=246$~GeV one finds  $\lambda_{h_0}\approx 0.02$, not outlandish but roughly an order of magnitude too low. However as stressed above, the breaking of flavour degeneracy in the Higgs VEVs, and a full treatment of runnings and thresholds will almost certainly disrupt the quartic parameters in the potential. Indeed the diagonal component $h_0$ must ultimately be replaced by a single dominant component.

Next let us consider the electroweak couplings. 
In the present example, with $N_f\approx  56$, we find that at the UV fixed point they have a value $\alpha^*_{EW} = 4\pi\tilde{\alpha}/N_f\approx 0.33$. They are required to be of order $1/30$ at the electroweak scale, and matching to 
this value fixes the appropriate trajectory in figure \ref{fig:flow}.
Indeed despite the large value of the `t Hooft-like coupling $\tilde{\alpha}$, the corresponding electroweak coupling will run to be  less in the  IR than the $SU(N)$ coupling $\alpha_g$, because the latter is in a Banks-Zaks like regime and runs exceedingly slowly, whereas the electroweak coupling runs rather rapidly. One may simply solve for its leading order running up from the value $1/30$ near the (extended)  Pati-Salam breaking  scale $M_{PS}$ (neglecting the contribution between $M_{PS}$ and the weak scale), to the energy scale $\mu_*$ 
where the $SU(2)_{L/R}$ couplings saturate their fixed point value. That is 
\begin{align}
{\tilde{\alpha}}(M_{PS})^{-1}-{\tilde{\alpha}}(\mu_*)^{-1}  &~\approx~  \frac{3}{4} (t_{PS}-t_*) ~ .
\end{align}
We may use this to estimate the energy at which the electroweak couplings saturate the fixed point values:
\begin{align}
{\mu_*} & ~\approx~ e^{90\pi /N_f}M_{PS} \nonumber \\
& ~\approx~ 150 M_{PS}~.
\end{align}
Note that this ratio of scales is determined entirely by the number of $SU(2)$ flavours $N_f$.
The scale $M_{PS}$ is then determined by the choice of $m_{\tilde Q}^2$ independently of the electroweak Higgs mass-squared parameter $m_0^2$. Below this scale the running of the entire theory reverts to that of the usual SM. 

\bigskip
\vspace{0.5cm}

\noindent {\bf Acknowledgements}: We are especially grateful to  Esben M\o lgaard for help with the beta functions. 
This work is partially supported by the Danish National Research Foundation under the grant DNRF:90.

\bibliographystyle{apsrev4-1}
%\bibliography{final}

\end{document}